\documentclass[prb,twocolumn,aps,showpacs,fixfloats]{revtex4}
\usepackage{graphicx}
\usepackage{subfigure}
\usepackage{float}
\usepackage{bm}
\usepackage{amsmath,amssymb}

\usepackage{latexsym}
\usepackage{color}
\usepackage{enumerate}
\usepackage{pdfpages}
\usepackage{tikz}
\usepackage{hyperref}

\begin{document}
\newcommand{\s}{\scriptscriptstyle}
\newcommand{\uu}{\uparrow \uparrow}
\newcommand{\ud}{\uparrow \downarrow}
\newcommand{\du}{\downarrow \uparrow}
\newcommand{\dd}{\downarrow \downarrow}
\newcommand{\ket}[1] { \left|{#1}\right> }
\newcommand{\bra}[1] { \left<{#1}\right| }
\newcommand{\bracket}[2] {\left< \left. {#1} \right| {#2} \right>}
\newcommand{\vc}[1] {\ensuremath {\bm {#1}}}
\newcommand{\tr}{\text{Tr}}
\newcommand{\Trans}{\ensuremath \Upsilon}
\newcommand{\Refl}{\ensuremath \mathcal{R}}

\title{ Lifshitz model in the presence of spin-orbit 
coupling }

\author
{M. E. Raikh}

\affiliation{ Department of Physics and
Astronomy, University of Utah, Salt Lake City, UT 84112}
\begin{abstract}
Wave function of a localized state created by a 
short-range impurity in two dimensions falls off with distance, $r$,
from the impurity as $\frac{1}{r^{1/2}}\exp(-\frac{r}{a})$, where $a$
is the localization radius.  With  randomly positioned {\em identical} impurities
with low concentration, $n \ll a^{-2}$, the level smears into a band
due to the overlap of the impurity wave functions. This is the essence 
of the Lifshitz model. We demonstrate that, upon incorporation of the spin-orbit coupling, the impurity wave functions acquire oscillating
factors which, subsequently, modify their overlap.
 As a result of such modification,
the density of states develops singularities at certain energies.

\end{abstract}

\maketitle

\section{Introduction}

Conventional Lifshitz model\cite{Lifshitz,Gredeskul} describes single-electron states in a system of randomly-positioned {\em identical} short-range 
impurities. It is assumed that each impurity creates a localized level with a radius of wave 
function, $a$, much smaller than a typical  distance between the neighboring impurities.
In two dimensions this distance is $n^{-1/2}$,
where $n$ is the impurity concentration. Due to
overlap of the wave functions of neighboring
impurities, individual levels smear into an impurity band. 
Presence of a small parameter $na^2\ll 1$ allows to establish 
the form of the density of states as a function of $E-E_0$,  
which is the electron energy measured from the level position, $E_0$.
The key argument summarized in the book Ref. \onlinecite{book} why this density can be found analytically is that the eigenstates
of the disordered system are composed of  pairs of impurities  which
hybridize the respective wave functions. As a result of this hybridization
one of the levels shifts up from $E_0$, while the other level shifts down.
Thus, the density of states in the Lifshitz model has the form of two peaks.

Assume now that the bare Hamiltonian of the 2D system contains a spin-orbit
term e.g. of the Rashba type.\cite{Rashba}  This certainly does not have an effect on the
density of states if the spin splitting of the bare band is much smaller 
than $E_0$. In the opposite limit,   the localized states
are classified according to {\em helicity}. Importantly, 
the wave functions of the short-range impurities get modified  dramatically by spin-orbit coupling compared to a
simple exponential decay.\cite{Galstyan,Chaplik,
Mkhitaryan,Voloshin,Hutchinson}
The main message of the present
paper is that in the limit of strong spin-orbit coupling the shape of the density of states in the Lifshitz model changes dramatically, namely, it develops {\em singularities} at certain energies.

\vspace{2mm}

\section{A single short-range impurity in the presence of spin-orbit coupling}

The density of states of free  electrons with a quadratic Hamiltonian
${\hat H}_0=\frac{\hbar^2{\bm k}^2}{2m}$, where $m$ is the effective mass and 
${\bm k}$ is the wave vector, is energy-independent.
Upon adding a spin-orbit term
\begin{equation}
\label{so}
{\hat H}_{SO}= \alpha[{\bm k}\times {\hat {\bm \sigma}}]\cdot{\bm n},
\end{equation}
where ${\bm \sigma}$ is the vector of the Pauli matrices, $\alpha$ is the spin-orbit
constant, and ${\bm n}$ is the normal to the 2D plane, to ${\hat H}_0$
modifies the density of states qualitatively.\cite{Galstyan,Chaplik}
Indeed, the spectrum of the full Hamiltonian ${\hat H}_0+{\hat H}_{SO}$ consists
of two branches
\begin{equation}
\label{branches}
E_{1,2}({\bm k})=\frac{\hbar^2k^2}{2m}\mp\alpha|{\bm k}|.
\end{equation}
The corresponding wave functions have the form
\begin{equation}
\label{wavefunctions}
\Psi_{{\bm k}}^{(1,2)}({\bm r})=e^{i{\bm kr}}\chi_{{\bm k}}^{(1,2)},
\end{equation}
where the spinors $\chi_{{\bm k}}^{(1,2)}$ are defined as
\begin{equation}
\label{chi}
\chi_{{\bm k}}^{(1)}=\frac{1}{\sqrt{2}}
\begin{pmatrix}
e^{i\Phi_{{\bm k}}}\\
-1
\end{pmatrix}, \hspace{3mm}
\chi_{{\bm k}}^{(2)}=\frac{1}{\sqrt{2}}
\begin{pmatrix}
1\\
e^{-i\Phi_{{\bm k}}}
\end{pmatrix}.
\nonumber
 \end{equation}
Here $\Phi_{{\bm k}}$ is the azimuthal angle of the vector ${\bm k}$.
It is crucial that the lower branch of the spectrum 
Eq.~(\ref{branches}) has a minimum at
\begin{equation}
\label{k0}
k=k_0= \frac{m\alpha}{\hbar^2}.
\end{equation}
Near this minimum the spectrum can be simplified as
\begin{equation}
 \label{minimum}
 E_1({\bm k})=-\Delta +\frac{\hbar^2(k-k_0)^2}{2m},
\end{equation}
with the depth of the minimum 
$\Delta=\frac{m\alpha^2}{2\hbar^2}=\frac{\hbar^2k_0^2}{2m}$.
As a result,  the wave function of a  localized state created  by a
point-like impurity is composed of the free-electron state with 
energies close to $-\Delta$. The density of these states behaves 
as $(E+\Delta)^{-1/2}$, i.e. it has a one-dimensional  character.\cite{Galstyan}

Calculation of the level position in this setting was carried out in Ref.~
\onlinecite{Chaplik}. We reproduce it below for pedagogical reasons and in order to
generalize it later to the impurity pairs.  

Denote with $U({\bm r})$ the impurity potential. The solution of the Schr{\"o}dinger equation 
\begin{equation}
\label{Schroedinger}
\Bigl[{\hat H_0} +{\hat H_{SO}}+ U({\bm r})\Bigr]\Psi({\bm r})
=E_0\Psi({\bm r})
\end{equation}
for the level, $E_0$, close to $-\Delta$
can be searched in the form of the combination of
the states of the lower branch only
\begin{equation}
\label{combination}
\Psi({\bm r})=\int d{\bm k}A({\bm k})\chi_{{\bm k}}^{(1) }e^{-i{\bm kr}}.
\end{equation}
Substitution of this form into Eq. (\ref{Schroedinger}) yields
\begin{equation}
\label{substitution}
\int d{\bm k}A({\bm k})\chi_{{\bm k}}^{(1) }e^{-i{\bm kr}}
\Bigl[E_1({\bm k})-E_0\Bigr] =-U({\bm r})\Psi(0).
\end{equation}
In Eq. (\ref{substitution}) we took into account that $U({\bm r})$
is short-ranged.
Multiplying this equation by $\chi_{{\bm q}}^{(1)*}e^{i{\bm qr}}$
and integrating $d{\bm r}$, we get

\begin{equation}
\label{A}
(2\pi)^2A({\bm q})\Bigl(E_1 ({\bm q})-E_0\Bigr)
=
-\Bigl(\Psi(0)\chi_{{\bm q}}^{(1)*}  \Bigr)
\int d{\bm r}U({\bm r})e^{i{\bm qr}}.
\end{equation}
Expressing $A({\bm q})$, substituting it into 
Eq. (\ref{combination}) and setting ${\bm r}=0$, we arrive to
the self-consistency equation
\begin{equation}
\label{consistency}
\Psi(0)=-\frac{1}{(2\pi)^2}\int d{\bm q}\frac{\chi_{{\bm q}}^{(1)}\Bigl(\Psi(0)\chi_{{\bm q}}^{(1)*}  \Bigr)}
{E_1({\bm q})-E_0}
\int d{\bm r}U({\bm r})e^{i{\bm qr}}.
\end{equation} 
This equation defines the position of the level, $E_0$.
For isotropic potential, the dependence on the direction
of ${\bm q}$ disappears from the integral $d{\bm r}$, yielding $J_0(qr)$, where $J_0$ is the Bessel function. Then
the angular integration over ${\bm q}$ can be easily performed. Concerning the integral over $|{\bm q}|$, it comes from the domain $\left(|{\bm q}|-k_0\right)\ll k_0$.
Finally, the solution of Eq. (\ref{consistency}) takes the
form\cite{Chaplik}
\begin{equation}
\label{degenerate}
\varepsilon_0=-\Delta-E_0=\frac{\pi^2mk_0^2}{2\hbar^2}
\Biggl[ \int\limits_0^\infty dr r U( r)J_0(k_0r)    \Biggr]^2.
\end{equation}
The solution is doubly degenerate with respect to the components of respective  spinors. The 
meaning of $\varepsilon_0$ is the binding energy
measured from the minimum of the spectrum of the
lower branch.

To estimate the binding energy predicted by 
Eq.  (\ref{degenerate}), we assume that the magnitude of
the potential is $U_0$, while the radius of potential is $a$. To insure that the wave function does not change within the interval
$r<a$ the condition $k_0a<<1$ should be met,
which is equivalent to the replacement of the 
Bessel function in the integrand by $1$.
Then, within a numerical factor, we get

\begin{equation}
\label{estimate}
\varepsilon_0\sim (k_0a)^2\frac{U_0^2}{\frac{\hbar^2}{ma^2}}.
\end{equation}
To test the assumptions made in course of solving  of Eq.~(\ref {consistency}) this binding energy must be much smaller
than the depth $\frac{\hbar^2k_0^2}{2m}$ of the minimum in the spectrum
(to justify the integration over ${\bm q}$). Also, this binding
energy should be much smaller than $U_0$ (to replace $\Psi({\bm r})$ by $\Psi (0)$.
The first requirement leads to the usual condition $U_0\ll 
\frac{\hbar^2}{ma^2}$. The second requirement yields a complimentary condition $(k_0a)^2U_0\ll\frac{\hbar^2}{ma^2} $, which is
weaker.
    
The form of the wave function   at distances
$r\gg a$ is established from Eqs. (\ref{combination}) and (\ref{A}). Introducing a wave vector
\begin{equation}
\label{kc} 
k_c=\left(\frac{2m\varepsilon_0}{\hbar^2}  \right)^{1/2}
\end{equation}
and the dimensionless variable $z$ defined as
\begin{equation}
\label{according}
z= \frac{q-k_0}{k_{c}},   
\end{equation}
and performing the angular integration, we get for 
a nonzero component of a spinor
\begin{equation}
\label{according}
\Psi(r)\propto \int\limits_0^\infty
\frac{J_0(k_0r+k_crz)}{z^2+1}.
\end{equation}
Since characteristic $z$ is $\sim 1$, the 
localization length of $\Psi (r)$ is $r_c\sim k_c^{-1}$.
For $r<r_c$, the $z$-dependence in
the argument of the Bessel function can be neglected, so that $\Psi(r) \propto J_0(k_0r)$. For $r>r_c$
the Bessel function can be replaced by a large-argument
asymptote: $J_0(z)\approx \left( \frac{2}{\pi z} \right)^{1/2}\cos\left(z-\frac{\pi}{4}   \right) $. This yields
\begin{equation}
\label{asymptote}
\Psi(r)\propto\frac{\cos\left(k_0r-\frac{\pi}{4}   \right)}{r^{1/2}}\exp\left(-k_c r   \right).
\end{equation}

We see that, unlike the conventional impurities,
the wave function, in addition to the exponential decay,  contains an oscillating factor with a period $2\pi/k_0$. This behavior
is illustrated in the figure.
In the next section we recalculate this oscillations into the
splitting of the levels of two impurities.


\section{Two impurities}

Let the impurities be located at $\pm \frac{1}{2}{\bm \rho}$, so that the net potential has the
form

\begin{equation}
\label{rho}
{\tilde U}\left({\bm r}\right)=U\left({\bm r}-\frac{\bm \rho}{2}\right)+
U\left({\bm r}+\frac{\bm \rho}{2}\right).
\end{equation}
It is straightforward to generalize Eq. (\ref{A}) to the
case of two impurities
\begin{align}
\label{B}
&(2\pi)^2A({\bm q})\Bigl(E_1 ({\bm q})-E_0\Bigr)= 
-\int d{\bm r}U({\bm r})e^{i{\bm qr}} 
\times\nonumber\\
&\Biggl\{ e^{i\frac{{\bm q}{\bm \rho}}{2}} 
\Bigl(\Psi\left(\frac{\bm \rho}{2}\right)\chi_{{\bm q}}^{(1)*}  \Bigr)
+e^{-i\frac{{\bm q}{\bm \rho}}{2}}
\Bigl(\Psi\left(-\frac{\bm \rho}{2} \right)\chi_{{\bm q}}^{(1)*}  \Bigr)\Biggr\}.              
\end{align}    
Expressing $A(\bm q)$ from Eq. (\ref{B}), substituting
it into Eq.~(\ref{combination}) and setting $\bm r=\frac{\bm \rho}{2}$ and $\bm r=-\frac{\bm \rho}{2}$,
we arrive at the system of equations for 
$\Psi\left(\frac{\bm \rho}{2}\right)$ and
$\Psi\left(-\frac{\bm \rho}{2}\right)$.

To cast this system in a concise form we take the
advantage of the fact that the distance, $\rho$,
between the impurities is much bigger than $a$,
so that the splitting of the levels due to
the overlap of the impurity wave functions is
smaller than $\varepsilon_0$ given by 
Eq. (\ref {degenerate}). We introduce the
following notations for the elements of the spinors   $\Psi\left(\frac{\bm \rho}{2}\right)$
and $\Psi\left(-\frac{\bm \rho}{2}\right)$

\begin{equation}
\label{spinors}
\Psi\left(\frac{\bm \rho}{2}\right)=
\begin{pmatrix}
 a_1\\
 b_1
\end{pmatrix}, \hspace{3mm}
\Psi\left(-\frac{\bm \rho}{2}\right)=
\begin{pmatrix}
a_2\\
b_2
\end{pmatrix}.
\nonumber
\end{equation}
Then the generalization of Eq. 
(\ref{consistency}) to the case of two impurities
takes the form
\begin{align}
\label{SYSTEM}
& \kappa a_1 =\bigl[J_0(k_0\rho)a_2 -iJ_1(k_0\rho)b_2\bigr]\exp\left(-k_c \rho   \right), \nonumber\\
& \kappa b_1 = \bigl[iJ_1(k_0\rho)a_2 +J_0(k_0\rho)b_2\bigr]\exp\left(-k_c \rho   \right), \nonumber\\
&\kappa a_2 =\bigl[J_0(k_0\rho)a_1 -iJ_1(k_0\rho)b_1\bigr]\exp\left(-k_c \rho   \right), \nonumber\\
& \kappa b_2 = \bigl[iJ_1(k_0\rho)a_1 +J_0(k_0\rho)b_1\bigr]\exp\left(-k_c \rho   \right),
\end {align}
where the parameter $\kappa$ is related to the
energy, $\varepsilon$, and to the single-impurity binding energy $\varepsilon_0$ as
\begin{equation}
\label{root}
 \kappa=
 \Bigl(\frac{\varepsilon}{\varepsilon_0}   \Bigr)^{1/2}
 -1.
\end{equation}

The four solutions of the
system Eq. (\ref{SYSTEM}) can be easily found, namely

\begin{align}
\label{KAPPA}
&\kappa^{+}(\rho)=\pm \bigl[J_0(k_0\rho) +J_1(k_0\rho)\bigr]
\exp\left(-k_c \rho   \right), \nonumber\\
&\kappa^{-}(\rho)=\pm \bigl[J_0(k_0\rho) -J_1(k_0\rho)\bigr]
\exp\left(-k_c \rho   \right).
\end{align}
Concerning the eigenvectors, their structure can be illustrated e.g. for positive $\kappa^{+}$ when it is symmetric,
namely, $a_2=a_1$ and $b_2=b_1=ia_1$.

Eq. (\ref{KAPPA}) allows to trace how the overlap-induced splitting of single-impurity
 levels evolves with the increase of the distance between the impurities. For 
 $k_0\rho \ll 1$ the Bessel function 
 $J_0(k_0\rho)$ should be replaced by $1$,
 while $J_1(k_0\rho)$ should be replaced by $0$.
 Then the splitting is of 1D-type and does not
 depend on the spin-orbit parameter, $k_0$.
 In the opposite limit $k_0\rho \gg 1$, substituting the large-argument asymptotes of the Bessel functions into Eq. (\ref{KAPPA}) yields

\begin{align}
\label{KAPPA1}
&\kappa^{+}(\rho)=\pm \frac{2\sin (k_0\rho)}{(k_0\rho)^{1/2}}
\exp\left(-k_c \rho   \right),
\nonumber\\
&\kappa^{-}(\rho)=\pm \frac{2\cos( k_0\rho)}{(k_0\rho)^{1/2}}
\exp\left(-k_c \rho   \right).
\end{align}
The $1/\rho^{1/2}$ behavior of the prefactor 
confirms  that, at  large $\rho$, the splitting is of the 2D-type and oscillates with distance
due to the spin-orbit coupling. Since $k_c \gg k_0$,
the splitting of the impurity levels strongly oscillates with distance, $\rho$, between the impurities. 

\vspace{3mm}

\section{Density of states}

Assume now that the density, $n$,  of the short-range impurities is finite. As a result of the overlap of 
single-impurity wave functions, a level with binding energy
$\varepsilon_0$ gets smeared into a band.



A basic  assumption of the conventional Lifshitz  model,\cite{Lifshitz} is that the width of the band
is much smaller than $\varepsilon_0$. This amounts to
the smallness of the parameter  
$nk_{c}^{-2} \ll 1$. For such low  densities the overlap-induced splitting Eq. (\ref{KAPPA}) of two levels  at a typical distance $\rho \sim n^{-1/2}$ contains an exponential factor, $\exp\left[-k_{c}n^{-1/2}\right]$, which is small by virtue of this parameter.

A simplified argument which allows to establish the shape of the density of states, $g(\varepsilon)$,  in the vicinity of $\varepsilon =\varepsilon_0$ goes as follows.\cite{book} A shift of a given single-impurity level will be anomalously small, if it has no neighboring impurities within a circle of a certain radius, $\rho_{\varepsilon }$. This
radius is anomalously large compared to the typical distance between the impurities, i.e. $\rho_{\varepsilon} \gg n^{-1/2}$.
With sub-logarithmic accuracy, the condition for $\rho_{\varepsilon }$ reads
\begin{equation}
\label{CONDITION}
\exp\left(-k_c\rho_{\varepsilon}\right)=\frac{|\varepsilon-\varepsilon_0|}{\varepsilon_0},
\end{equation} 
where the right-hand side is $\kappa$ defined by Eq. (\ref{root}) and taken at 
$|\varepsilon-\varepsilon_0|\ll \varepsilon_0$. Probability that the neighboring impurities are absent in the circle with a
radius  $\rho_{\varepsilon}$ is equal to 
$\exp\left(-\pi n \rho_{\varepsilon}^2   \right)$. Substitution of  
$\rho_{\varepsilon}$ found from Eq. (\ref{CONDITION}) into this probability, 
yields the exponent in the density of states
\begin{equation}
\label{exponent}
g(\varepsilon)\propto \exp\Bigg[-\frac{\pi n}{k_c^2}\ln^2\Big(\frac{\varepsilon_0}{|\varepsilon-\varepsilon_0|}      \Big)    \Bigg].
\end{equation}
Since the ratio $\frac{n}{k_c^2}$ is small,
Eq. (\ref{exponent}) describes a sharp minimum 
in a parametrically narrow domain
\begin{equation}
\label{narrow}
|\varepsilon -\varepsilon_0| \sim 
\varepsilon_0\exp\Big[-\frac{k_c}{(\pi n)^{1/2}}      \Big].
\end{equation}

To incorporate the spin-orbit coupling, we use the modified splitting Eq. (\ref{KAPPA1}) and present the
density of states as a sum 
$g(\varepsilon)=g^+(\varepsilon)+g^-(\varepsilon) $,
where $g^+$ and $g^-$ are defined as

\begin{align}
\label{pm}
&g^+(\varepsilon)=\int\limits_0^\infty d{\rho}F(\rho)
\delta\Bigg[\frac{|\varepsilon-\varepsilon_0|}{2\varepsilon_0}-\frac{4|\sin k_0\rho|}{(k_0\rho)^{1/2}}   e^{-(k_c\rho   )}   \Bigg], \\
&g^-(\varepsilon)=\int\limits_0^\infty d{\rho}F(\rho)
\delta\Bigg[\frac{|\varepsilon-\varepsilon_0|}{2\varepsilon_0}-\frac{4|\cos k_0\rho|}{(k_0\rho)^{1/2}}   e^{-(k_c\rho   )}   \Bigg].
\end{align}
Here $F(\rho)$ is the nearest neighbor distribution
\begin{equation}
\label{NN}
F(\rho)=2\pi n\rho\exp\left(-\pi n\rho^2   \right).
\end{equation}
Our key point is that, due to the spin-orbit coupling,
the level splitting defined by $\kappa^+$
turns to zero at certain distances,
$\rho_N=\frac{\pi N}{k_0}$, while $\kappa^-$ 
turns to zero at $\rho_{N+\frac{1}{2}}=\frac{\pi (N+\frac{1}{2})}{k_0}$. Substituting 
\begin{equation}
\label{DELTA}    
\rho=\rho_N+\delta \rho_N
\end{equation}
into the first equation of Eq. (\ref{pm}), we present 
$g^+(\varepsilon)$ as a sum
\begin{align}
\label{SUM} 
&g^+(\varepsilon)=\sum_{N}\frac{(\pi N)^{1/2}e^{k_c\rho_N}}
{4}F(\rho_N)\nonumber\\
&\times \int d\delta\rho_N \delta\Bigg[ \frac{|\varepsilon-\varepsilon_0|}{8\varepsilon_0}
(\pi N)^{1/2}e^{k_c\rho_N}  -|\sin (k_0\delta\rho_N)|  \Bigg].
\end{align}
In deriving Eq. (\ref{SUM}) we made two assumptions: 
$\delta \rho_N \ll \rho_N$ and $k_c\delta \rho_N \ll 1$.
As it is seen from Eq. (\ref{SUM}), characteristic $\delta \rho_N$ is $\sim k_0^{-1}$, so that the first assumption is valid for $N\gg 1$, while the second condition is ensured by the relation $k_c \ll k_0$, which was assumed above.
 
 The "memory" about the spin-orbit coupling in Eq. (\ref{SUM}) is 
encoded in the combination $\sin (k_0\delta\rho_N)$. Taking the limit $k_0\rightarrow 0$ and replacing the summation 
over $N$ by integration, one recovers the result Eq. (\ref{exponent}). 

To reveal the role of spin-orbit coupling we
perform the integration in Eq. (\ref{SUM}) with a help of the $\delta$-function, we find
\begin{equation}
\label{FINAL}
g^+(\varepsilon)=\frac{1}{4}\sum_{N}\frac{F(\rho_N)}
{\Big[\frac{\exp{(-2k_c\rho_N)}}{\pi N} -\Big(\frac{\varepsilon -\varepsilon_0}{8\varepsilon_0}     \Big)^2        \Big]^{1/2}       }.
\end{equation}
The expression for $g^-(\varepsilon)$ has a
similar form with replacement of $N$ by
$N+\frac{1}{2}$.
Note that each term in Eq. (\ref{FINAL}) exhibits a {\em square-root
singularity} near energies

\begin{equation}
\label{singular}
\varepsilon_N^\pm=\varepsilon_0\Bigg[   1\pm \frac{8}{(\pi N)^{1/2}}
\exp{\Big(-\frac{2\pi k_c}{k_0}N     \Big )   }\Bigg].
\end{equation}
On the other hand, by virtue of the small parameter $\frac{k_c}{k_0}$, the intervals $\{\varepsilon_N^+,\varepsilon_N^-\} $ for different $N$ overlap. Thus, the overall shape of the density of states is smooth. To uncover the role of the discreteness of $N$, we transform Eq.~ (\ref{FINAL}) using the Poisson summation

\begin{equation}
\label{summation}
\sum_NS(N)=\int dx S(x)+2\sum_l\int dx S(x)
\cos(2\pi lx).
\end{equation}
In our case, the function $S(x)$ has a form
\begin{equation}
\label{form} 
S(x)=\frac{1}{4}\frac{F\left(\frac{\pi x}{k_0}\right)}
{\Big[\frac{\exp{\left(-\frac{2\pi k_cx}{k_0}\right)}}{\pi x} -\Big(\frac{\varepsilon -\varepsilon_0}{8\varepsilon_0}     \Big)^2        \Big]^{1/2}       }.
\end{equation}
 
The denominator in Eq. (\ref{form}) turns to zero
at 
\begin{equation} 
x=x_{\varepsilon}\approx \frac{k_0}{2\pi k_c}
\ln \Bigl(\frac{8\varepsilon_0}{\varepsilon-\varepsilon_0}   \Bigr)^2.
\end{equation}
The first term in Eq. (\ref{summation})
reproduces the standard result Eq. (\ref{exponent})
for the density of states in the absence of spin-orbit coupling. To evaluate the terms with $l \ge 1$ one should expand the integrand around $x_{\varepsilon}$.
This leads to the oscillating spin-orbit component in the density
of states $\propto \cos(2\pi lx_{\varepsilon})$.
\vspace{3mm}

\section{Concluding remarks}
The prime observation of the present paper is that, in the presence of  spin-orbit coupling, 
the pairs of impurities located at certain distances from each other {\em do not} hybridize their levels. This leads to singularities in the 
in the density of states at certain energies.

There is a certain similarity between the impurity state in a 2D electron gas with strong spin-orbit coupling and the in-gap state created by a magnetic impurity
in a superconductor.\cite{Yu,Shiba,Rusinov,Glazman}
Similarly to Eq. (\ref{asymptote}), the wave function
of the in-gap state oscillates with distance as
$\cos k_Fr$, where $k_F$ is the Fermi momentum.
However, the limit of low density of magnetic impurities considered in the present paper is not
relevant for superconductors. Rather, in the relevant regime, there are many impurities within the localization radius determined by the coherence length. In this limit, the self-consistent Born approximation applies, and the two-peak structure of
the density of the impurity states transforms into a
semicircle.\cite{Glazman}

\end{document}